\documentclass[12pt]{article}

\usepackage{epsfig,amsfonts,amsthm}
\usepackage{amsmath,amssymb}

\newcommand{\be}{\begin{equation}}
\newcommand{\ee}{\end{equation}}
\newcommand{\bea}{\begin{eqnarray}}
\newcommand{\eea}{\end{eqnarray}}
\newcommand{\bs}{\begin{subequations}}
\newcommand{\es}{\end{subequations}}
\newcommand{\no}{\nonumber\\}

\newcommand{\Z}{\mathbb{Z}}
\newcommand{\Yukawa}[9]{\left( \begin{array}{ccc}
#1 & #2 & #3 \\ #4 & #5 & #6 \\ #7 & #8 & #9 \end{array} \right)}

\def\lsim{\mathrel{\rlap{\lower4pt\hbox{\hskip1pt$\sim$}}
    \raise1pt\hbox{$<$}}}         
\def\gsim{\mathrel{\rlap{\lower4pt\hbox{\hskip1pt$\sim$}}
    \raise1pt\hbox{$>$}}}         

\title{
\normalsize \hfill CFTP/15-011 \\*[7mm]
\LARGE $SO(10)$ models with flavour symmetries: \\
Classification and examples}
\author{I.~P.~Ivanov\footnote{E-mail: {\tt igor.ivanov@tecnico.ulisboa.pt}}
\ {\small and} 
L.~Lavoura\footnote{E-mail: {\tt balio@cftp.tecnico.ulisboa.pt}} 
\\
{\small CFTP, Instituto Superior T\'ecnico, Universidade de Lisboa,
1049-001 Lisboa, Portugal}
\\
}

\date{\today}

\begin{document}
\maketitle
\begin{abstract}
We establish the full list of flavour symmetry groups which may be enforced,
without producing any further accidental symmetry,
on the Yukawa-coupling matrices of an $SO(10)$ Grand Unified Theory
with arbitrary numbers of scalar multiplets in the $\mathbf{10}$,
$\overline{\mathbf{126}}$,
and $\mathbf{120}$ representations of $SO(10)$.
For each of the possible discrete non-Abelian symmetry groups,
we present examples of minimal models
which do not run into obvious contradiction
with the phenomenological fermion masses and mixings.
\end{abstract}

\section{Introduction}

There is a long history of attempts
at explaining the fermion masses and mixings
through (discrete) symmetry groups
in models beyond the Standard Model (bSM).
They started,
back in the 1970's,
with guesses or hopes that permutation groups
might help explain and predict the patterns
of the quark masses and mixing~\cite{quark-old},
and over the course of decades evolved into
an elaborate group-theoretic machinery,
especially for the lepton sector---for recent reviews
see~\cite{fermions-review}.
In the simplest approach,
one assumes that several scalars exist
which couple to the fermions through Yukawa matrices
which inherit symmetries from the model bSM.
Consider,
for example,
the quark sector with the following Yukawa Lagrangian
with $n_\phi$ scalar doublets and $n_g$ generations:
\be
{\cal L}_Y = - \sum_{a=1}^{n_H}\, \sum_{i, j = 1}^{n_g} \bar Q_{L i} \left(
\Gamma^{a}_{ij} \phi_{a} d_{R j}
+ \Delta^{a}_{ij} \tilde \phi_{a} u_{R j} \right)
+ \mathrm{H.c.}
\label{yukawa-general}
\ee
If this Lagrangian inherits some symmetry from a high-energy model bSM,
then the Yukawa matrices $\Gamma^{a}_{ij}$ and $\Delta^{a}_{ij}$
are invariant under a transformation acting simultaneously
in the flavour spaces of the left-handed quark doublets $Q_{Li}$,
right-handed up-type quark singlets $u_{Rj}$,
down-type quark singlets $d_{Rj}$,
and scalar doublets $\phi_a$.
When the scalars acquire vacuum expectation values (vevs),
those Yukawa matrices produce the mass matrices
$M_d = \sum_a \Gamma^{a} \! \left\langle \phi_a^0 \right\rangle$ 
and $M_u  = \sum_a \Delta^{a} \! \left\langle \phi_a^0 \right\rangle^\ast$
and the symmetry may get lost.
However,
if the symmetry was \textit{ab initio}\/ sufficiently restrictive,
then the mass matrices might still have predictive power.
In this field of research one wants to use a symmetry group
to construct a model that is able to fit the known observables,
\textit{viz.}~the fermion masses and the mixing parameters,
without requiring fine-tuning and that is sufficiently predictive.
This activity naturally splits into two parts:
firstly,
to find which symmetry groups are available for a given model bSM,
and secondly,
to check which symmetry groups lead to masses and to mixing patterns
in agreement with the phenomenology.
In this paper we address only the first task.

Grand Unified Theories (GUTs) based on the group $SO(10)$
are particularly attractive in this context,
because in those theories
all the left- and right-handed fermions of each generation are united
in a single irreducible representation (irrep) $\mathbf{16}$ of $SO(10)$.
As a consequence,
all the Yukawa couplings take the simple form $f^T \Gamma_a H_a f$,
where $f$ stands for the column vector of the three fermionic $\mathbf{16}$
and the $\Gamma_a$ are $3 \times 3$ Yukawa-coupling matrices in family space.
The scalar multiplets $H_a$ may be
either $\mathbf{10}$ or $\overline{\mathbf{126}}$ of $SO(10)$,
which couple to the symmetric part of the tensor product
$\mathbf{16} \otimes \mathbf{16}$,
or $\mathbf{120}$ of $SO(10)$,
which couple to the antisymmetric part of the tensor product;
correspondingly,
the $\Gamma_a$ will be either symmetric or antisymmetric,
respectively.
The (anti)symmetry properties of the $\Gamma_a$
are preserved in weak-basis changes
\be
\label{weakbasis}
\Gamma_a \to \sum_b U_{ab} \left( W^T \Gamma_b W \right),
\ee
where $U$ is a unitary matrix which mixes the various scalar $\mathbf{10}$
(or $\overline{\mathbf{126}}$,
or $\mathbf{120}$)
and $W$ is a $3 \times 3$ unitary matrix which mixes the three
fermionic $\mathbf{16}$.
For an overview of flavour model-building opportunities with $SO(10)$, 
see the classical review~\cite{slansky-review}
and the more recent summary~\cite{senjanovic}.
In this paper we want to analyze which symmetry groups one may impose
on the Yukawa-coupling matrices of an $SO(10)$ GUT.
In our search,
we do not restrict ourselves in our choice of scalars---we derive results
that are valid for arbitrary numbers
of scalar multiplets $\mathbf{10}$,
$\overline{\mathbf{126}}$,
and $\mathbf{120}$ of $SO(10)$.
Thus,
we go far beyond not only the early $SO(10)$ models,
but also,
for instance,
the very recent study~\cite{Ferreira:2015jpa};
the examples presented in that paper emerge 
as specific cases of our general classification.

It might happen that by imposing a symmetry group $G$
one ends up producing a model which is symmetric
under a larger group $G' \supset G$.
(This is sometimes called an `accidental symmetry'.)
A common instance of this occurs when $G$ is a cyclic group and $G'$ is $U(1)$.
In our analysis,
we shall always try to identify accidental symmetries which may be present
in the Yukawa-coupling matrices that we write down.

In our search for discrete,
non-Abelian symmetry groups,
we shall use the method of~\cite{IvanovVdovin2013}.
Namely,
we shall firstly derive all the possible Abelian symmetry groups;
then,
we shall use group-theoretical methods
to combine the Abelian symmetry groups in all possible ways
into non-Abelian groups.
Here,
knowing the full list of possible Abelian symmetry groups,
\textit{i.e.}~knowning that no other Abelian group may be a subgroup
of the non-Abelian symmetry group that we are constructing,
is a strong factor limiting the possible choices.

It is important to stress that in this paper we only focus
on the Yukawa-coupling sector of the $SO(10)$ GUT.
We disregard the scalar sector,
\textit{viz.}\ the scalar potential,
of the GUT.
This sector depends,
in particular,
on which scalar $SO(10)$ irreps exist beyond the $\mathbf{10}$,
$\overline{\mathbf{126}}$,
and $\mathbf{120}$;
on whether those scalar representations constitute basic building blocks
of a model or they are just the effective combination of other scalar irreps;
on whether the GUT is supersymmetric or not;
and on whether the scalar potential is renormalizable or not.
Depending on the potential,
some symmetries that exist in the Yukawa couplings
may or may not be partially broken.
So,
the accidental symmetries that may be present
in the Yukawa couplings that we write down
might be broken in the scalar potential,
but we shall not deal on that issue here.

This paper is organized as follows.
In section~\ref{class} we provide and derive the full list of symmetries
that may occur in the Yukawa couplings of an $SO(10)$ GUT.
In section~\ref{examples} we give the simplest models
that realize each of the \emph{discrete non-Abelian}\/ symmetries
that we have listed in section~\ref{class}.
We summarize our findings in section~\ref{conclusions}.

\section{Full classification of the symmetries} \label{class}

In this section we shall list all the symmetry groups,
both discrete and continuous, 
which can be used in flavour model building in $SO(10)$ GUTs
with an arbitrary number of Higgs multiplets in the irreps $\mathbf{10}$,
$\overline{\mathbf{126}}$,
and $\mathbf{120}$.
The end result of this section is the following list of groups:
\bs
\label{full-list}
\bea
\mbox{discrete Abelian:} & &
\Z_2, \quad
\Z_3, \quad
\Z_4, \quad
\Z_2\times\Z_2;
\label{full-list-1}
\\
\mbox{continuous Abelian:} & &
U(1), \quad
U(1) \times \Z_2, \quad
U(1) \times U(1);
\label{full-list-2}
\\
\mbox{discrete non-Abelian:} & &
S_3, \quad
D_4, \quad
Q_4, \quad
A_4, \quad
S_4, \quad
\Delta(54) \left/ \ \Z_3^\mathrm{center} \right., \quad
\Sigma(36); \hspace*{6mm}
\label{full-list-3}
\\
\mbox{continuous non-Abelian:} & &
O(2), \quad
O(2) \times U(1), \quad
\left[ U(1) \times U(1) \right] \rtimes S_3,
\no & &
SU(2), \quad
SU(2) \times U(1), \quad
SO(3), \quad
SU(3).
\label{full-list-4}
\eea
\es
In~(\ref{full-list-3}),
\be
\label{Z3center}
\Z_3^\mathrm{center} = \left\{
\mbox{diag} \left( 1,\ 1,\ 1 \right),\
\mbox{diag} \left( \omega,\ \omega,\ \omega \right),\
\mbox{diag} \left( \omega^2,\ \omega^2,\ \omega^2 \right)
\right\},
\quad \omega \equiv \exp{\left( 2 i \pi / 3 \right)}
\ee
is the center of $SU(3)$.

Our claim is that
trying to enforce any symmetry group which is not in the list~(\ref{full-list})
unavoidably produces a model whose full symmetry group,
including the accidental symmetries,
is in the list.

We derive the classification~(\ref{full-list})
by using the methods developed in~\cite{IvanovKeusVdovin2012}
and~\cite{IvanovVdovin2013},
\textit{viz.}\ we firstly identify all possible Abelian symmetry groups
and we then construct the non-Abelian groups
as extensions of the Abelian ones.

Readers who are not interested in the detailed derivation
of~(\ref{full-list}) may skip this section.

\subsection{Rephasing symmetries}

We start with symmetries which act on fermion families and on scalars
just through rephasings.
In analogy with~(\ref{weakbasis}),
one has
\be
\label{jvity}
\Gamma_a \to e^{i\psi_a} S^T \Gamma_a S = \Gamma_a,
\quad \mbox{where} \
S = \mathrm{diag} \left (e^{i\alpha_1},\ e^{i\alpha_2},\ e^{i\alpha_3} \right).
\ee
The phases $\psi_a$ arise
from the transformation of the scalar multiplets $H_a$
which couple through the matrices $\Gamma_a$;
the phases $\alpha_{1,2,3}$ refer to
the transformation of the three fermion families.
The transformation~(\ref{jvity}) does not mix matrices $\Gamma_a$
corresponding to scalars in different irreps of $SO(10)$,
\textit{i.e.}\ it does not mix the $\Gamma_a$ linked to $H_a$
in the $\mathbf{10}$ with those linked to $H_a$
in either the $\overline{\mathbf{126}}$ or the $\mathbf{120}$.

We consider the following problem:
With $n_S$ symmetric $\Gamma_a$ and $n_A$ antisymmetric $\Gamma_a$,
which rephasing symmetry groups can one have?\,\footnote{Clearly,
\label{generalu1}
the Yukawa interactions $f^T H_a f$ are invariant
under the simultaneous global rephasing
of all the fermions $f$ by a phase $\delta$
and of all the Higgs multiplets $H_a$ by a phase $- 2 \delta$.
We only look for symmetries above and beyond
this trivial global rephasing invariance.}
This problem can be systematically solved
through the Smith Normal Form (SNF) technique
explained in~\cite{IvanovKeusVdovin2012,IvanovNishi2013}.
Adapting it to the problem at hand,
we write the equations
\be
\sum_l d_{kl}\, \alpha_l \equiv \alpha_i + \alpha_j + \psi_a = 2 \pi n_k. 
\label{SNF-1}
\ee
Equation~(\ref{SNF-1}) states that the phase $\alpha_i + \alpha_j + \psi_a$
acquired by a {\em nonzero}\/ entry $\left( \Gamma_a \right)_{ij}$
must be an integer multiple of $2 \pi$.
To this end,
we have introduced indices $k$ that refer to all the nonzero entries
of any of the Yukawa-coupling matrices.
We have moreover represented all the phases,
including the $n_S + n_A$ phases $\psi_a$,
by $\alpha_l$,
where $l = 1, 2, 3, 4, \ldots, \left( 3 + n_S + n_A \right)$.
The integer-valued coefficients $d_{kl}$ may take the values either 0 or 1 or 2.

Equations~(\ref{SNF-1}) constitute 
a system of $m$ linear equations
for the $3 + n_S + n_A$ phases $\alpha_l$,
where $m$ is the total number of independent\footnote{Since the $\Gamma_a$
are either symmetric or antisymmetric,
not all their off-diagonal matrix elements are independent.}
nonzero entries of all the $\Gamma_a$.
We must now analyze the properties of the matrix $D = \{d_{kl}\}$;
namely,
we must find its SNF,
read out its diagonal values,
and from them write the corresponding symmetry group.
This procedure is described in more detail
in~\cite{IvanovKeusVdovin2012,IvanovNishi2013}.

\subsubsection{Single matrix $\Gamma$}

Let us suppose that there is a single matrix $\Gamma_a$.
There are then only four phases $\alpha_l$,
with $l = 1,2,3,4$ and $\alpha_4 = \psi_a$;
moreover,
in every row of $D$ the last entry is always $1$.
The first three entries of each row of $D$ may be,
up to permutations,
either $(2,\, 0,\, 0)$ or $(1,\, 1,\, 0)$.
For instance,
a row $(2,\, 0,\, 0,\, 1)$ of $D$ corresponds to nonzero $\Gamma_{11}$;
a row $(1,\, 0,\, 1,\, 1)$ of $D$ corresponds to nonzero
$\Gamma_{13}$ and $\Gamma_{31}$
(remember that
all the matrices $\Gamma_a$ are either symmetric or antisymmetric).

The number of possible matrices $\Gamma$ is small,
so they can be checked one by one.
For example,
\be
\label{duitp}
\mbox{if} \
\Gamma \sim \Yukawa{\times}{0}{0}{0}{\times}{0}{0}{0}{\times},
\quad \mbox{then} \
D = \left(\begin{array}{cccc}
2 & 0 & 0 & 1\\
0 & 2 & 0 & 1\\
0 & 0 & 2 & 1
\end{array}\right).
\ee
(In the matrix $\Gamma$ in~(\ref{duitp}),
and below,
a $\times$ represents a nonzero entry.)
Through simple manipulations\footnote{The allowed manipulations are:
effecting permutations of the order of the rows and/or columns of $D$;
flipping the signs of any rows and/or columns of $D$;
and adding any row or column of $D$ to any other row or column.}
of the matrix $D$ in~(\ref{duitp}),
one arrives at its SNF,
which is 
\be
D_\mathrm{SNF} = \left(\begin{array}{cccc}
1 & 0 & 0 & 0\\
0 & 2 & 0 & 0\\
0 & 0 & 2 & 0
\end{array}\right).
\label{uvipt}
\ee
The crucial point is the following:
the manipulations of $D$ which bring it to its SNF
leave invariant the set of solutions of the system~(\ref{SNF-1}).
Namely,
that system is transformed into a new one,
\be
\sum_l \left( D_\mathrm{SNF} \right)_{kl}\, \tilde \alpha_l = 2 \pi \tilde n_k.
\label{uvisr}
\ee
Thus,
the SNF in~(\ref{uvipt}) yields
\be
\tilde \alpha_1 = 2 \pi \tilde n_1, \quad
\tilde \alpha_2 = \pi \tilde n_2, \quad
\tilde \alpha_3 = \pi \tilde n_3,
\ee
which means that
the relevant symmetry group is $\Z_2 \times \Z_2$,
because both $\tilde \alpha_2$ and $\tilde \alpha_3$
are integer multiples of $\pi$.
The last column of $D_\mathrm{SNF}$ in~(\ref{uvipt}) is composed of zeros
and therefore places no restriction on $\tilde \alpha_4$;
this free $\tilde \alpha_4$ constitutes a $U(1)$ invariance
that exists for \emph{any}\/ Yukawa matrices
and represents the global rephasing mentioned in footnote~\ref{generalu1}.

Another example is
\be
\Gamma \sim \Yukawa{\times}{\times}{0}{\times}{\times}{0}{0}{0}{\times}
\ \Rightarrow \
D = \left(\begin{array}{cccc}
2 & 0 & 0 & 1\\
0 & 2 & 0 & 1\\
0 & 0 & 2 & 1\\
1 & 1 & 0 & 1
\end{array}\right)
\ \Rightarrow \
D_\mathrm{SNF} = \left(\begin{array}{cccc}
1 & 0 & 0 & 0\\
0 & 1 & 0 & 0\\
0 & 0 & 2 & 0\\
0 & 0 & 0 & 0
\end{array} \right)
\ \Rightarrow \ G = \Z_2.
\ee
One sees that adding one off-diagonal nonzero entry
to the $\Gamma$ of~(\ref{duitp}) reduces the symmetry group
from $\Z_2 \times \Z_2$ to $\Z_2$.

If there are less than three nonzero entries in $\Gamma$,
then the system~(\ref{uvisr}) is unable to fix
all three $\tilde \alpha_{1,2,3}$;
this implies a symmetry group which contains $U(1)$ factors.
For instance,
\be
\Gamma \sim \Yukawa{\times}{0}{0}{0}{0}{\times}{0}{\times}{0}
\ \Rightarrow \
D = \left(\begin{array}{cccc}
2 & 0 & 0 & 1\\
0 & 1 & 1 & 1
\end{array}\right)
\ \Rightarrow \
D_\mathrm{SNF} = \left(\begin{array}{cccc}
1 & 0 & 0 & 0\\
0 & 1 & 0 & 0
\end{array}\right)
\ \Rightarrow \ G = U(1),
\label{bjprt}
\ee
since the SNF in~(\ref{bjprt}) yields
no constraint on $\tilde \alpha_3$.

By examining all the possible matrices $\Gamma$ in this way,
we arrive at the list~(\ref{uxipr}) of possible symmetries.
\bs
\label{uxipr}
\bea
\mbox{For a symmetric $\Gamma$:} & & U(1)\times U(1),
\quad U(1)\times \Z_2,
\quad U(1),
\quad \Z_2 \times \Z_2,
\quad \Z_2.
\\
\mbox{For an antisymmetric $\Gamma$:} & & U(1) \times U(1),
\quad U(1).
\eea
\es
The difference between symmetric and antisymmetric matrices $\Gamma$
arises because for antisymmetric $\Gamma$ there are less possibilities
for nonzero matrix elements---only off-diagonal matrix elements
may be nonzero.
The explicit matrices corresponding to each symmetry group are,
up to permutations,
\bs
\label{euipt}
\bea
U(1) \times U(1):
& \!\! \!\! \!\! &
\left( \begin{array}{ccc}
\times & 0 & 0 \\ 0 & 0 & 0 \\ 0 & 0 & 0
\end{array} \right),
\quad
\left( \begin{array}{ccc}
0 & \times & 0 \\ \times & 0 & 0 \\ 0 & 0 & 0
\end{array} \right);
\label{U1U1-type}\\[2mm]
U(1) \times \Z_2: & \!\! \!\! \!\! &
\Yukawa{\times}{0}{0}{0}{\times}{0}{0}{0}{0};
\label{U1Z2-type}
\\[2mm]
U(1):
& \!\! \!\! \!\! &
\Yukawa{\times}{\times}{0}{\times}{0}{0}{0}{0}{0}
\subset
\Yukawa{\times}{\times}{0}{\times}{\times}{0}{0}{0}{0},
\quad
\Yukawa{\times}{0}{0}{0}{0}{\times}{0}{\times}{0},
\quad
\Yukawa{0}{\times}{\times}{\times}{0}{0}{\times}{0}{0};
\hspace*{9mm}
\label{U1-type}
\\[2mm]
\Z_2\times \Z_2: & \!\! \!\! \!\! &
\Yukawa{\times}{0}{0}{0}{\times}{0}{0}{0}{\times};
\label{Z2Z2-type}
\\[2mm]
\Z_2: & \!\! \!\! \!\! &
\Yukawa{\times}{\times}{0}{\times}{0}{0}{0}{0}{\times}
\subset
\Yukawa{\times}{\times}{0}{\times}{\times}{0}{0}{0}{\times}.
\label{Z2-type}
\eea
\es
Any other matrices $\Gamma$---except those
which are permutations of one of the matrices
in~(\ref{euipt})---possess no symmetry at all.

\subsubsection{Several matrices $\Gamma_a$}

If there are several Yukawa-coupling matrices $\Gamma_a$,
then each row of the matrix $D$ has,
up to permutations,
one of the following forms
\bs
\bea
&(2,\, 0,\, 0\, |\, 0,\, \ldots,\, 1,\, \ldots,\, 0), & \label{structure-1}
\\
&(1,\, 1,\, 0\, |\, 0,\, \ldots,\, 1,\, \ldots,\, 0). & \label{structure-2}
\eea
\es
(The form~(\ref{structure-1}) is available only for symmetric $\Gamma_a$.)
This allows for more possibilities than the single-$\Gamma_a$ case.
For example,
if there are two matrices
\be
\label{beiot}
\Gamma_1 \sim \Yukawa{\times}{0}{0}{0}{0}{\times}{0}{\times}{0}
\quad \mbox{and} \quad
\Gamma_2 \sim \Yukawa{0}{0}{0}{0}{\times}{0}{0}{0}{\times},
\ee
then
\be
D =
\left(\begin{array}{ccc|cc}
2 & 0 & 0 & 1 & 0\\
0 & 1 & 1 & 1 & 0\\
0 & 2 & 0 & 0 & 1\\
0 & 0 & 2 & 0 & 1
\end{array}\right)
\ \Rightarrow \
D_\mathrm{SNF} = \left(\begin{array}{ccccc}
1 & 0 & 0 & 0 & 0\\
0 & 1 & 0 & 0 & 0\\
0 & 0 & 1 & 0 & 0\\
0 & 0 & 0 & 4 &0 
\end{array}\right)
\ \Rightarrow \ G = \Z_4.
\ee
The problem that one now faces
is how to efficiently check all possible combinations
of many matrices $\Gamma_a$.
In the following we make several observations that simplify,
and eventually allow one to solve,
that problem.

\paragraph{First observation:} If {\em the same}\/ nonzero entry
is present in both $\Gamma_a$ and $\Gamma_{a'}$,
then the two scalar multiplets $H_a$ and $H_{a'}$
must transform in the same way under the symmetry,
\textit{viz.}\ $\psi_a = \psi_{a^\prime}$.
But then,
the structures $\Gamma_a$ and $\Gamma_{a'}$ must completely coincide.
Thus,
any two matrices $\Gamma_a$ and $\Gamma_{a'}$
either do not have nonzero entries in the same position,
or they have nonzero entries
at fully identical positions.\footnote{This result does not depend on whether
the matrices $\Gamma_a$ and $\Gamma_{a'}$ are symmetric or antisymmetric;
if one of them is antisymmetric,
then it just does not borrow the nonzero diagonal entries
from the symmetric one.}

For example,
the symmetry group of the two matrices
\be
\Gamma_1 \sim \Yukawa{0}{\times}{0}{\times}{0}{\times}{0}{\times}{0}
\quad \mbox{and} \quad
\Gamma_2 \sim \Yukawa{0}{\times}{\times}{\times}{0}{0}{\times}{0}{0}
\ee
is the same as the symmetry group
of the merged matrix\footnote{This symmetry group happens to be trivial,
as we have seen in the previous subsection.
In group-theoretic terms,
the two $U(1)$ symmetry groups of $\Gamma_1$ and $\Gamma_2$
do not have a non-trivial intersection.}
\be
\Gamma_3 \sim \Yukawa{0}{\times}{\times}{\times}{0}{\times}{\times}{\times}{0}.
\ee

Thus,
the matrices $\Gamma_a$ may be grouped
in several sets of matrices with identical nonzero matrix elements,
and any two sets of matrices do not have any common
nonzero matrix element.\footnote{Having several Yukawa-coupling matrices
with the same texture will in general complicate the analysis
of the resulting mass matrices;
yet,
the {\em group-theoretic properties}---they are what we care about
here---are not sensitive to a proliferation of identical matrices.}

\paragraph{Second observation:} Matrices $\Gamma_a$
with a single entry\footnote{Whenever we talk
of a single-entry $\Gamma_a$,
we always have in mind only the \emph{independent nonzero}\/ entries
of that $\Gamma_a$.}
do not modify the symmetry group in any way.
This is because,
by adjusting the $\psi_a$ which transforms the scalar field $H_a$,
any single-entry $\Gamma_a$ will be symmetric
under any rephasing of the fermion fields that one wishes.

Therefore,
the rows of the matrix $D$ corresponding to a single-entry $\Gamma_a$
may safely be eliminated from $D$;
simultaneously,
the column of $D$ corresponding to the phase $\psi_a$ should also be removed.
One only needs to check matrices with more than one independent nonzero entry.

\paragraph{Third observation:} The matrix $D$ has at most six rows,
corresponding to the six possible nonzero entries in the matrices $\Gamma_a$.
As we have seen in the first observation,
the rows may be grouped into a few non-intersecting sets.
According to the second observation,
none of the sets is allowed to have just one row.
There are only five ways of grouping at most six rows in several sets,
when none of the sets has just one row:
\be
4+2, \quad 3+3, \quad 2+2+2, \quad 3+2, \quad 2+2.
\label{grouping}
\ee

\paragraph{Fourth observation:} The 2-,
3-,
and 4-entry matrices with non-trivial symmetry groups
have already been given
in~(\ref{U1Z2-type})--(\ref{Z2-type}):\,\footnote{Remember that
any matrix $\Gamma$ which is not in~(\ref{euipt})
does not possess any symmetry.}
\bs
\bea
\mbox{4-entry}: & &
\Yukawa{\times}{\times}{0}{\times}{\times}{0}{0}{0}{\times},
\Yukawa{\times}{0}{\times}{0}{\times}{0}{\times}{0}{\times},
\Yukawa{\times}{0}{0}{0}{\times}{\times}{0}{\times}{\times}
\quad (\mbox{symmetry} \ \Z_2);
\label{i1} \\
\mbox{3-entry}: & &
\Yukawa{\times}{\times}{0}{\times}{\times}{0}{0}{0}{0},
\Yukawa{\times}{0}{\times}{0}{0}{0}{\times}{0}{\times},
\Yukawa{0}{0}{0}{0}{\times}{\times}{0}{\times}{\times}
\quad (\mbox{symmetry} \ U(1));
\label{i2} \\ & &
\Yukawa{\times}{0}{0}{0}{\times}{0}{0}{0}{\times}
\quad (\mbox{symmetry} \ \Z_2 \times \Z_2);
\label{i3} \\
\mbox{2-entry}: & &
\Yukawa{\times}{0}{0}{0}{0}{\times}{0}{\times}{0},
\Yukawa{0}{\times}{0}{\times}{0}{0}{0}{0}{\times},
\Yukawa{0}{0}{\times}{0}{\times}{0}{\times}{0}{0}
\quad (\mbox{symmetry} \ U(1));
\label{i4} \\ & &
\Yukawa{\times}{0}{0}{0}{\times}{0}{0}{0}{0},
\Yukawa{0}{0}{0}{0}{\times}{0}{0}{0}{\times},
\Yukawa{\times}{0}{0}{0}{0}{0}{0}{0}{\times}
\quad (\mbox{symmetry} \ U(1) \times\Z_2); \hspace*{7mm}
\label{i5} \\ & &
\Yukawa{0}{\times}{\times}{\times}{0}{0}{\times}{0}{0},
\Yukawa{0}{0}{\times}{0}{0}{\times}{\times}{\times}{0},
\Yukawa{0}{\times}{0}{\times}{0}{\times}{0}{\times}{0}
\quad (\mbox{symmetry} \ U(1)).
\label{i6}
\eea
\es
The symmetry group of a collection of such matrices
is the intersection of the symmetry groups of the individual matrices.
Therefore,
in order to find symmetry groups beyond the ones already listed
in~(\ref{U1Z2-type})--(\ref{Z2-type}),
we only need to intersect the $U(1)$ groups
coming from matrices of the types~(\ref{i2}),
(\ref{i4}),
(\ref{i5}),
and~(\ref{i6}).

\bigskip

We are left with very few possible combinations
of non-intersecting 3- or 2-entry matrices
in combinations of the types $3+2$,
$2+2$,
and $2+2+2$.
We must check those combinations one by one.

The only
(but for permutations of the rows and columns)
combination of the type $3+2$ is
\be
\Gamma_1 \sim \Yukawa{\times}{\times}{0}{\times}{\times}{0}{0}{0}{0},
\quad
\Gamma_2 \sim \Yukawa{0}{0}{\times}{0}{0}{\times}{\times}{\times}{0}.
\ee
One easily sees that the corresponding symmetry group is $U(1)$.

There are only two possible combinations of the type $2+2+2$:
\bs
\bea
&
\Gamma_1 \sim \Yukawa{\times}{0}{0}{0}{\times}{0}{0}{0}{0},
\quad 
\Gamma_2 \sim \Yukawa{0}{\times}{0}{\times}{0}{0}{0}{0}{\times},
\quad  
\Gamma_3 \sim \Yukawa{0}{0}{\times}{0}{0}{\times}{\times}{\times}{0};
&
\label{Z2-structurex}
\\
&
\Gamma_1 \sim \Yukawa{\times}{0}{0}{0}{0}{\times}{0}{\times}{0},
\quad
\Gamma_2 \sim \Yukawa{0}{0}{\times}{0}{\times}{0}{\times}{0}{0},
\quad
\Gamma_3 \sim \Yukawa{0}{\times}{0}{\times}{0}{0}{0}{0}{\times}.
&
\label{Z3-structures}
\eea
\es
The corresponding symmetry groups are $\Z_2$ and $\Z_3$,
respectively.

There are several possible combinations of the type $2+2$.
Some of them are subsets of~(\ref{Z3-structures})
and have the same $\Z_3$ symmetry group.
The combination~(\ref{beiot})
has symmetry group $\Z_4$,
as we have already seen.
All the other $2 + 2$ combinations give symmetry groups
which are either $U(1)$ or $\Z_2$.

\bigskip

We have thus arrived at the conclusion that
Yukawa-coupling matrices in $SO(10)$ models
can only have the Abelian symmetries in~(\ref{euipt}),
plus $\Z_3$ and $\Z_4$.
No other group can be the {\em full}\/ rephasing-symmetry group
of any collection of either symmetric or antisymmetric Yukawa matrices.

\subsection{$\Z_3 \times \Z_3$} \label{Z3Z3}

In the previous section we have assumed that
any Abelian symmetry acts through rephasing
in both the fermion and scalar sectors.
We now relax this assumption and consider symmetry transformations
which act through rephasing in the fermion sector
and through an {\em  arbitrary unitary matrix}\/ in the scalar sector.
Any single transformation
can be brought to this form through an appropriate basis change
of the fermion generations.
We want to discover whether symmetry groups of this type exist.

In this case,
the transformation~(\ref{jvity}) must be generalized to
\be
\label{gamma-mixing}
\Gamma_a \to S^T \Gamma_a S = \sum_b v_{ab} \Gamma_b,
\ee
with coefficients $v_{ab}$ forming
a unitary transformation matrix $V = \{ v_{ab} \}$
in the space of the scalars $H_a$.
Since $V$ is unitary,
it has eigenvectors.
By performing a basis change in the space of the $H_a$ 
we can arrive at matrices $\bar \Gamma_a$
which are the eigenvectors of $V$;
the eigenvalues have modulus~1 because $V$ is unitary,
\textit{i.e.}\ they are phases $e^{i\psi_a}$.
In this way we reduce~(\ref{gamma-mixing}) to~(\ref{jvity}).

Thus,
although the condition~(\ref{gamma-mixing}) appears to offer more freedom, 
that freedom in reality corresponds only
to a meaningless change of basis in scalar space. 
The available rephasing-symmetry groups
are exactly the same as before,
\textit{viz.}~$\Z_3$,
$\Z_4$,
and the groups in~(\ref{euipt}).

Still,
we must remember that an overall rephasing of the scalar fields
just corresponds to the trivial $U(1)$ transformation
mentioned in footnote~\ref{generalu1}.
Therefore,
we must consider the action of symmetry transformations up to such a rephasing.
This means looking not just for Abelian symmetry groups
belonging to either $U(3)$ or $SU(3)$,
but also for Abelian groups belonging to
$PSU(3) \simeq U(3) \left/ \, U(1)^\mathrm{center} \right.
\simeq SU(3) \left/ \, \Z_3^\mathrm{center} \right.$,
where
\be
U(1)^\mathrm{center} =
\left\{ \mathrm{diag} \left( e^{i \theta},\
e^{i \theta},\ e^{i \theta} \right) \right\}
\ee
is the center of $U(3)$
and $\Z_3^\mathrm{center}$ is the center of $SU(3)$.
It turns out that this allows for
\emph{only one}\footnote{The proof of this fact
can be found in~\cite{IvanovKeusVdovin2012}.}
further Abelian group:
the group $\Z_3 \times \Z_3 = \Delta(27) \left/ \ \Z_3^\mathrm{center} \right.$.

The non-Abelian group $\Delta(27)$ is the subgroup of $SU(3)$
generated by the matrices
\bea
\label{a3-generator}
A_3 &=& \Yukawa{1}{0}{0}{0}{\omega}{0}{0}{0}{\omega^2},
\\
\label{D}
D &=& \Yukawa{0}{e^{i \psi_1}}{0}{0}{0}{e^{i \psi_2}}
{e^{- i \left( \psi_1 + \psi_2 \right)}}{0}{0},
\eea
where the phases $\psi_1$ and $\psi_2$ are arbitrary.
This group contains the center of $SU(3)$,
\textit{viz.}~(\ref{Z3center}).
It is easy to check that the factor group of $\Delta(27)$
by its center $\Z_3^\mathrm{center}$ is the Abelian group $\Z_3 \times \Z_3$.
This Abelian group cannot be represented just through a rephasing;
its faithful irreducible representation is not one-dimensional
but rather three-dimensional.

Let us look for Yukawa-coupling matrices
transforming as a representation of $\Z_3 \times \Z_3$.
We firstly identify matrices $\Gamma_{1,2,3}$ such that
\be
A_3 \Gamma_1 A_3 = \Gamma_1, \quad
A_3 \Gamma_2 A_3 = \omega^2 \Gamma_2, \quad
A_3 \Gamma_3 A_3 = \omega \Gamma_3.
\ee
We find
\be
\label{S3-structures}
\Gamma_1 = \Yukawa{f_1}{0}{0}{0}{0}{g_1}{0}{g_1}{0}, \quad 
\Gamma_2 = \Yukawa{0}{0}{g_2}{0}{f_2}{0}{g_2}{0}{0}, \quad  
\Gamma_3 = \Yukawa{0}{g_3}{0}{g_3}{0}{0}{0}{0}{f_3}.
\ee
We then enforce $D$-invariance of
$\left\{ \Gamma_1,\ \Gamma_2,\ \Gamma_3 \right\}$.
One easily finds that
\be
D^T \Gamma_1 D = \Gamma_2, \quad
D^T \Gamma_2 D = \Gamma_3, \quad
D^T \Gamma_3 D = \Gamma_1,
\ee
provided
\be
\label{sziut}
f_2 = f_1\, e^{2 i \psi_1}, \quad
g_2 = g_1\, e^{- i \psi_1}, \quad
f_3 = f_2\, e^{2 i \psi_2}, \quad
g_3 = g_2\, e^{- i \psi_2}.
\ee
Thus,
the matrices~(\ref{S3-structures}) are $\Z_3 \times \Z_3$-invariant
provided phases $\psi_1$ and $\psi_2$ exist such that~(\ref{sziut}) apply.

We have thus found that $\Z_3 \times \Z_3$
is another possible Abelian symmetry of $SO(10)$ Yukawa-coupling matrices.
The full list of possible Abelian symmetries is thus
\bs
\bea
\label{bvuit}
& & \Z_2, \quad \Z_3, \quad \Z_4, \quad
\Z_2 \times \Z_2, \quad \Z_3 \times \Z_3,
\\
& & U(1), \quad U(1) \times \Z_2, \quad U(1) \times U(1).
\eea
\es

It so happens that the set of matrices $\Gamma_{1,2,3}$ in~(\ref{S3-structures})
with the proviso~(\ref{sziut})
is accidentally invariant under a larger symmetry group.
Let us define
\be
\label{b-generator}
B = \left( \begin{array}{ccc}
-1 & 0 & 0 \\ 0 & 0 & - e^{i \psi_2} \\ 0 & - e^{- i \psi_2} & 0
\end{array} \right).
\ee
Then,
the matrices~(\ref{S3-structures}) which satisfy~(\ref{sziut})
also satisfy
\be
B^T \Gamma_1 B = \Gamma_1, \quad
B^T \Gamma_2 B = \Gamma_3, \quad
B^T \Gamma_3 B = \Gamma_2.
\ee
Thus,
$\left\{ \Gamma_1,\ \Gamma_2,\ \Gamma_3 \right\}$
is automatically $B$-invariant
and its symmetry group is not just $\Z_3 \times \Z_3$:
it is actually $\Delta(54) \left/ \ \Z_3^\mathrm{center} \right.$,
where $\Delta(54)$ is the subgroup of $SU(3)$ generated by $A_3$,
$D$,
and $B$.
For this reason,
the group $\Z_3 \times \Z_3$ does not appear in~(\ref{full-list-1});
rather,
$\Delta(54) \left/ \ \Z_3^\mathrm{center} \right.$ is in~(\ref{full-list-3}).

\subsection{Discrete non-Abelian symmetries}

We next want to find the discrete \emph{non-Abelian}\/ symmetry groups
that may be used in the Yukawa sector of $SO(10)$ models.
Our analysis follows closely~\cite{IvanovVdovin2013},
where the analogous problem was solved
for the scalar sector of the three-Higgs-doublet model.
Indeed,
our results are exactly the same as in~\cite{IvanovVdovin2013};
we therefore repeat only briefly the argument in that paper.

Any non-Abelian {\em discrete}\/ group $G$
contains (usually many) Abelian subgroups $A$.
We must firstly have the {\em full}\/ list
of all possible discrete Abelian groups $A$.
We already know that list to be~(\ref{bvuit}).
Thus,
we want to know which non-Abelian discrete groups exist
which only have Abelian subgroups in~(\ref{bvuit}).

We note that all the groups in~(\ref{bvuit}) have group orders
with prime factors 2 and 3 only.
Therefore,
by Cauchy's lemma,
the order of any non-Abelian group
which only has Abelian subgroups in~(\ref{bvuit})
must be of the form $2^a 3^b$.
Now,
according to Burnside's theorem,
any group with order $2^a 3^b$ contains a normal Abelian subgroup.
The fact that we are looking for subgroups of $PSU(3)$
allows one to derive a stronger conclusion~\cite{IvanovVdovin2013}:
$G$ contains a normal \emph{maximal}\/ Abelian subgroup.
Let now $A$ denote that subgroup.
Then,
the group $G$ has structure
\be
G = A\, .\, K,\quad \mbox{where} \quad K \subseteq \mbox{Aut}(A),
\ee
\textit{i.e.}~the group $G$ is constructed as an extension of $A$
through a subgroup of the automorphism group of $A$.
Since we already have the full list~(\ref{bvuit}) of possible $A$,
we have to
\begin{enumerate}
\item find their automorphism groups $\mbox{Aut} (A)$,
\item find all the subgroups $K$ of the automorphism groups,
\item for each pair $A$ and $K$,
construct all the extensions of $A$ through $K$.
\end{enumerate}
At the end we will still need to check whether the resulting models
have not acquired any accidental symmetries,
especially continuous ones.
We leave that task to section~\ref{examples}.

We now follow the steps above for each of the groups in~(\ref{bvuit}):
\begin{enumerate}
\item $\mbox{Aut} \left( \Z_2 \right) = \{ e \}$,\footnote{The symbol $e$
denotes the identity transformation.}
hence no non-Abelian extension of $\Z_2$ is possible.
\item $\mbox{Aut} \left( \Z_3 \right) = \Z_2$,
therefore the only possible non-Abelian extension of $\Z_3$ is
$\Z_3 \rtimes \Z_2 = S_3$.
\item $\mbox{Aut} \left( \Z_4 \right) = \Z_2$,
therefore there are two possible non-Abelian extensions of $\Z_4$:
$\Z_4 \rtimes \Z_2 = D_4$ and $\Z_4\, .\, \Z_2 = Q_4$.
\item $\mbox{Aut} \left( \Z_2\times \Z_2 \right) = S_3$.
The group $S_3$ has subgroups $\Z_2$,
$\Z_3$,
and $S_3$.
Therefore,
the possible non-Abelian extensions of $\Z_2 \times \Z_2$ are
$\left( \Z_2 \times \Z_2 \right) \rtimes \Z_2 = D_4$,
$\left( \Z_2 \times \Z_2 \right) \rtimes \Z_3 = A_4$,
and $\left( \Z_2 \times \Z_2 \right) \rtimes S_3 = S_4$.
\item $\mbox{Aut} \left( \Z_3\times \Z_3 \right) = GL(2,3)$.
This is the group of general linear transformations
of a two-dimensional space over the finite field $\mathbb{F}_3$,
\textit{i.e.}\ the group of invertible $2 \times 2$ matrices
with matrix elements which are integers modulo~3.
The group $GL(2,3)$
has order 48 and has group elements of order 2,
3,
4,
and 6~\cite{groupprops}.
It turns out,
however,
that combining an element of order 3 with
$\Z_3 \times \Z_3$ always leads to a continuous symmetry.
Therefore,
only two choices for discrete extensions of $\Z_3 \times \Z_3$ remain:
\bs
\label{biptp}
\bea
\left( \Z_3 \times \Z_3 \right) \rtimes \Z_2
&=& \Delta(54) \left/ \ \Z_3^\mathrm{center} \right.,
\\
\left( \Z_3 \times \Z_3 \right) \rtimes \Z_4
&=& \Sigma \left( 36 \right).
\eea
\es
The two groups~(\ref{biptp}) are subgroups of $PSU(3)$ of order~18 and~36,
respectively.
Their preimages in $SU(3)$ are $\Delta(54)$ and $\Sigma(36\phi)$,
respectively,
which have order~54 and~108,
respectively.
\end{enumerate}

We have thus finished the derivation of~(\ref{full-list-3}).

\subsection{Continuous non-Abelian groups}

When studying rephasing symmetries,
we have identified three continuous Abelian groups:
$U(1)$,
$U(1) \times Z_2$,
and $U(1)\times U(1)$.
We now want to see how they can be extended
to non-Abelian groups.\footnote{No non-Abelian symmetry group
of the form $U(1) \times G$,
where $G$ is discrete non-Abelian,
may exist,
because $G$ would necessarily have only $\Z_2$ Abelian subgroups
and no such non-Abelian $G$ exists.}

We start by a more accurate description of $U(1)$ subgroups
when we pass from $SU(3)$ to $PSU(3)$.
There are two types of $U(1)$'s in $SU(3)$.
The first one is parameterized as
\be
\label{u11}
U(1)_1: \quad \mathrm{diag} \left(1,\ e^{i\alpha},\ e^{-i\alpha} \right),
\quad \alpha \in \left[ 0, 2\pi \right),
\ee
and,
provided $\alpha \neq 0$ and $\alpha \neq \pi$,
it has three distinct eigenvalues.
The second type,
$U(1)_2$,
is parameterized
$\mathrm{diag} \left( e^{2i\alpha},\ e^{-i\alpha},\ e^{-i\alpha} \right)$
and,
provided $\alpha \neq 0$ and $\alpha \neq \pm 2 \pi / 3$,
it has a twice-degenerate eigenvalue.
Because of this difference in the dimensionality of their subspaces,
the two groups $U(1)_1$ and $U(1)_2$ cannot be mapped onto each other 
by any basis transformation of $SU(3)$.
The center of $SU(3)$ is in $U(1)_2$ but not in $U(1)_1$.
We factor it out by defining
\be
\label{u12}
U(1)_2: \quad
\mathrm{diag}
\left( e^{2 i \alpha / 3},\ e^{-i \alpha / 3},\ e^{-i \alpha / 3} \right),
\quad \alpha \in \left[ 0, 2\pi \right). 
\ee
With this definition,
$U(1)_1$ and $U(1)_2$ both belong to $PSU(3)$,
they only intersect at the unit matrix,
and they serve as basis vectors on the torus of rephasing transformations.
The matrices~(\ref{U1-type})
are invariant under the two different $U(1)$s as
\be
\label{U11U12-type}
U(1)_1:\quad 
\Yukawa{\times}{0}{0}{0}{0}{\times}{0}{\times}{0};
\quad
U(1)_2:\quad 
\Yukawa{0}{0}{0}{0}{\times}{\times}{0}{\times}{\times},
\quad
\Yukawa{0}{\times}{\times}{\times}{0}{0}{\times}{0}{0}.
\ee
We must extend the two $U(1)$'s to non-Abelian groups separately.

Both $U(1)_1$ and $U(1)_2$ are generated
by rephasing transformations $S_\alpha$,
\textit{viz.}~(\ref{u11}) and~(\ref{u12}).
Let us suppose that there is another symmetry $R$ of the model. 
We want to know what options there are
for the enlarged group $G = \left\langle R,\ S_\alpha \right\rangle$.
We consider the transformation $S^R_\alpha = R^{-1}S_\alpha R$.
This is also a symmetry of the model.
There are two options:
(A) either $S^R_\alpha$ is in the original $U(1)$,
and then it is equal to some $S_\beta$,
or (B) it is not in the original $U(1)$.
In case A,
the invertible transformation $R$ induces a group automorphism;
the group of automorphisms of $U(1)$ is just $\Z_2$.
Thus,
case~A subdivides into two:
either $\beta=\alpha$ (case A1) or $\beta=-\alpha$ (case A2).
Let us now examine in turn each of the three cases~A1,
A2,
and~B.

In case~A1,
$R$ commutes with $S_\alpha$.
Therefore,
$G = \left\langle R,\ S_\alpha \right\rangle$ is Abelian;
if it is larger than the original $U(1)$,
then we already know that it can only be
either $U(1)\times \Z_2$ or $U(1)\times U(1)$.

In case~A2 we obtain the non-Abelian group $U(1) \rtimes \Z_2 \simeq O(2)$.
This is possible only when $U(1)$ is $U(1)_1$,
\textit{cf.}~(\ref{u11}),
and the $\Z_2$ transformation
is the permutation of second and third generations.
It is not possible to build a group $U(1)_2 \rtimes \Z_2$,
because there is no unitary transformation capable of mapping
\be
\mathrm{diag}
\left( e^{2 i \alpha / 3},\ e^{-i \alpha / 3},\ e^{-i \alpha / 3} \right)
\quad \mathrm{into} \quad
\mathrm{diag}
\left( e^{- 2 i \alpha / 3},\ e^{i \alpha / 3},\ e^{i \alpha / 3} \right);
\ee
such a transformation would have to be antiunitary.
Therefore,
$O(2) \simeq U(1)_1 \rtimes \Z_2$
can be further enlarged to $O(2) \times U(1)_2$.
In fact,
\emph{any}\/ single-entry matrix is invariant under that group.

In case~B,
$S^R_\alpha$ does not belong to the initial $U(1)$.
Therefore,
it defines a different $U(1)$ subgroup of the full symmetry group.
Let us now look at the group algebra rather than at the group itself.
The generators of $S_\alpha$ and of $S^R_\alpha$,
which we denote $t$ and $t'$,
respectively,
define a two-dimensional subspace in the entire (8+1)-dimensional space
spanned by the generators of $u(3)$.
If $t$ and $t'$ commute,
then we once again have a $U(1) \times U(1)$ symmetry group;
this is possible only when $R$ acts by permutation.
In this way we can obtain $\left[ U(1)_1 \times U(2)_2 \right] \rtimes S_3$,
which is the symmetry group of three single-entry Yukawa-coupling matrices
with equal entries.

If $t$ and $t'$ do not commute,
then we must close their subalgebra by including other generators.
There exist very few subalgebras of $su(3)$,
and they lead to the following non-Abelian groups:
$SU(2)$ and $SO(3)$
(which have the same algebra)
and $SU(2) \times U(1)$.
In this way we finish the derivation of~(\ref{full-list-4}).

\section{Minimal models with discrete non-Abelian symmetry} \label{examples}

In the previous section
we have already written down the Yukawa-coupling matrices
for models with Abelian symmetries.
Those matrices may in general,
as we have pointed out,
be accompanied by an arbitrarily large number of single-entry matrices,
which do not alter the symmetry group in any way
because their intrinsic Abelian symmetry group
always is the most general possible,
\textit{viz.}\ $U(1)_1 \times U(1)_2$.

Unfortunately,
models with just an Abelian symmetry
typically have a rather large number of free parameters.
In this section we want to reduce this large freedom
by looking for minimal models with {\em non-Abelian}\/ symmetries.
Specifically,
we shall look for models with \emph{discrete}\/ symmetries,
since the ones with continuous symmetries
are in general much too restricted.

\subsection{Models based on $S_3$}

The group $S_3$ is generated by two transformations $t_3$ and $t_2$ such that
$\left( t_3 \right)^3 = \left( t_2 \right)^2 = e$
(the identity transformation)
and $t_2 t_3 t_2 = \left( t_3 \right)^{-1}$.
In a triplet representation of the generators,
we may make a basis transformation in family space
such that $t_3 \to A_3$,
where $A_3$ is the matrix~(\ref{a3-generator}).
Then,
$t_2 \to B$,
where $B$ is the matrix~(\ref{b-generator}),
which contains an arbitrary phase $\psi_2$.\footnote{In this paper
we always adopt symmetry generators with determinant $+1$,
\textit{viz.}~belonging to $SU(3)$.}

We have already seen that a $\Z_3$-invariant model
contains the three Yukawa-coupling matrices~(\ref{S3-structures}).
In order to extend $\Z_3$ to $S_3$
one must enforce invariance under $B$ of the set of those three matrices.
The matrix $\Gamma_1$ is $B$-invariant by itself alone,
while $B^T \Gamma_2 B = e^{i \delta} \Gamma_3,
B^T \Gamma_3 B = e^{- i \delta} \Gamma_2$ provided
\be
\label{bjiot}
f_2\, e^{i \left( 2\psi_2 - \delta \right)} = f_3,
\quad
g_2\, e^{- i \left( \psi_2 + \delta \right)} = g_3.
\ee
Thus,
the set $\left\{ \Gamma_2,\ \Gamma_3 \right\}$ is $S_3$-invariant
if the conditions~(\ref{bjiot}) are satisfied.
Since the phases $\psi_2$ and $\delta$ are arbitrary,
those conditions simply translate into
\be
\left| f_2 \right| = \left| f_3 \right|, \quad
\left| g_2 \right| = \left| g_3 \right|.
\label{S3-conditions}
\ee
One may change the relative phase between the second and third fermion families
in order to change $\psi_2$.
One may also rephase the Higgs multiplets and thereby change $\delta$.
In this way one may,
for instance,
achieve $f_2 = f_3$ and $g_2 = g_3$.
We still have some rephasing freedom to set,
for instance,
both $f_1$ and $f_2 = f_3$ real
while $g_1$ and $g_2 = g_3$ remain complex;
or,
alternatively,
to set $f_1$ and $g_1$ real
while $f_2 = f_3$ and $g_2 = g_3$ remain complex.
Anyway,
there are six degrees of freedom in $\Gamma_{1,2,3}$.
(When the Higgs fields acquire vevs,
additional degrees of freedom appear.)

One may ask whether some of the matrix elements in $\Gamma_{1,2,3}$ may be zero.
It turns out that,
if either $f_2$ or $g_2$ is zero,
then the symmetry group is promoted to $O(2)$.
On the other hand,
either $f_1$ or $g_1$ may be zero without leading to an enhanced symmetry.
The case $f_1 = g_1 = 0$,
\textit{i.e.}~$\Gamma_1 = 0$,
is possible from the group-theoretical point of view,
but it will make the charged-lepton and down-type-quark mass matrices
proportional to each other,
which is phenomenologically ruled out.

Another possibility is to assume that $\Gamma_1$ is antisymmetric. 
Then $f_1 = 0$ and the off-diagonal matrix elements are $g_1$ and $-g_1$.
This is possible if the scalar multiplet $H_1$
transforms as a $\mathbf{1}'$ of $S_3$,
since then $B^T \Gamma_1 B = - \Gamma_1$.

The full list of models with $S_3$ symmetry
and having no more than three Higgs multiplets
is given in Table~\ref{table-S3models}.
For model~1 and model~2 the Yukawa-coupling matrices are
\be
\Gamma_1 = \Yukawa{f_1}{0}{0}{0}{0}{g_1}{0}{g_1}{0}, \quad 
\Gamma_2 = \Yukawa{0}{0}{g_2}{0}{f_2}{0}{g_2}{0}{0}, \quad  
\Gamma_3 = \Yukawa{0}{g_2}{0}{g_2}{0}{0}{0}{0}{f_2},
\ee
with (for instance) real $f_1$ and $f_2$ and complex $g_1$ and $g_2$.
For model~3 and model~4,
\be
\Gamma_1 = \Yukawa{0}{0}{0}{0}{0}{g_1}{0}{-g_1}{0}, \quad 
\Gamma_2 = \Yukawa{0}{0}{g_2}{0}{f_2}{0}{g_2}{0}{0}, \quad  
\Gamma_3 = \Yukawa{0}{g_2}{0}{g_2}{0}{0}{0}{0}{f_2},
\ee
with real $g_1$ and $g_2$ and complex $f_2$.
\begin{table}[ht!]
\centering
\begin{tabular}{c|cc}
& $H_1$ & $\left( H_2, H_3 \right)$
\\ \hline \\*[-4mm]
model~1 &
$\left( \overline{\mathbf{126}}, \mathbf{1} \right)$ &
$\left( \mathbf{10}, \mathbf{2} \right)$
\\ \hline \\*[-4mm]
model~2 &
$\left( \mathbf{10}, \mathbf{1} \right)$ &
$\left( \overline{\mathbf{126}}, \mathbf{2} \right)$
\\ \hline \\*[-4mm]
model~3 &
$\left( \mathbf{120}, \mathbf{1}' \right)$ &
$\left( \mathbf{10}, \mathbf{2} \right)$
\\ \hline \\*[-4mm]
model~4 &
$\left( \mathbf{120}, \mathbf{1}' \right)$ &
$\left( \overline{\mathbf{126}}, \mathbf{2} \right)$
\\ \hline
\end{tabular}
\caption{Minimal $SO(10)$ models with symmetry $S_3$.
In each parenthesis,
the first number denotes the $SO(10)$ irrep
and the second number denotes the $S_3$ irrep.}
\label{table-S3models}
\end{table}

\subsection{Models based on $D_4$}

There are two ways to construct $D_4$ as an extension of an Abelian group:
$D_4 = \Z_4 \rtimes \Z_2$
and $D_4 = \left( \Z_2 \times \Z_2 \right) \rtimes \Z_2$.

\subsubsection{$D_4 = \Z_4 \rtimes \Z_2$} \label{d4}

This group is generated by two transformations $t_4$ and $t_2$ such that
$\left( t_4 \right)^4 = \left( t_2 \right)^2 = e$
and $t_2 t_4 t_2 = \left( t_4 \right)^{-1}$.
As before,
in a triplet representation one can make $t_4 \to A_4$ diagonal
through an appropriate basis change:
\be
A_4 = \mathrm{diag} \left( 1,\ i,\ -i \right).
\ee
Then,
$t_2 \to B$ with the matrix $B$ in~(\ref{b-generator}).

We next write down the Yukawa matrices~(\ref{beiot}),
which define the group $\Z_4$:
\be
\label{D4-model}
\Gamma_1 = \Yukawa{f_1}{0}{0}{0}{0}{g_1}{0}{g_1}{0},
\quad
\Gamma_2 = \Yukawa{0}{0}{0}{0}{f_2}{0}{0}{0}{f_3}.
\ee
Clearly,
$A_4 \Gamma_1 A_4 = \Gamma_1$ and $A_4 \Gamma_2 A_4 = - \Gamma_2$.
We require that $\left\{\Gamma_1,\ \Gamma_2 \right\}$ be invariant under $B$.
We find that $B^T \Gamma_1 B = \Gamma_1$ automatically,
but imposing $B^T \Gamma_2 B = \sigma \Gamma_2$
is only possible when $\sigma = \pm 1$
and
\be
f_2\, e^{2 i \psi_2} = \sigma f_3.
\ee
This implies $\left| f_2 \right| = \left| f_3 \right|$.
The relative phase between $f_2$ and $f_3$ may be offset by $\psi_2$,
making $f_2 = f_3$.
Finally,
one may use the fermion and Higgs rephasing freedom
to make $f_1$ and $f_2 = f_3$ real,
while $g_1$ remains complex.

The above minimal $D_4$ model
is built with two non-equivalent $D_4$ singlets,
\textit{viz.}\ both $\Gamma_1$ and $\Gamma_2$ transform into themselves
under either $A_4$ or $B$.
There are four different singlets of $D_4$,
denoted $\mathbf{1}_{pq}$,
where the subscripts $p = \pm 1$ and $q = \pm 1$ reflect the actions
of $A_4$ and $B$,
respectively.
They correspond to the following matrices:
\be
\label{D4-singlets}
\Gamma^{(D_4)}_{++} = \Yukawa{f}{0}{0}{0}{0}{g}{0}{g}{0}, \quad
\Gamma^{(D_4)}_{+-} = \Yukawa{0}{0}{0}{0}{0}{h}{0}{-h}{0}, \quad
\Gamma^{(D_4)}_{-\pm} = \Yukawa{0}{0}{0}{0}{l}{0}{0}{0}{\pm l\, e^{2 i \psi_2}}.
\ee
Combining these,
one can construct several models with two distinct singlets.
Most of those models have an accidental continuous symmetry;
the only choice for which the full symmetry group remains $D_4$
and is not accidentally augmented to a continuous symmetry
is precisely $(\mathbf{1}_{++},\, \mathbf{1}_{-\pm})$,
\textit{i.e.}\ the matrices~(\ref{D4-model}) with $|f_2| = |f_3|$.

We may add more $D_4$ singlets,
either in the same or in different one-dimensional irreps of $D_4$,
and this leads to several more non-equivalent models.
Unfortunately,
all of them share the problem that
their mass matrices have a block-diagonal form
which leads to the decoupling of the first generation from the other two,
producing CKM and PMNS matrices in disagreement with the phenomenology.

These problems can be avoided in a minimal way by using one $\mathbf{1}_{++}$
and one doublet of $D_4$;
the corresponding models are shown in Table~\ref{table-D4models-2}.
The corresponding Yukawa-coupling matrices are
\be
\label{vvipt}
\Gamma_1 = \Yukawa{f_1}{0}{0}{0}{0}{g_1}{0}{g_1}{0}, \quad
\Gamma_2 = \Yukawa{0}{g_2}{0}{\pm g_2}{0}{0}{0}{0}{0}, \quad
\Gamma_3 = \Yukawa{0}{0}{g_2}{0}{0}{0}{\pm g_2}{0}{0},
\ee
where the plus sign holds for models~5 and~6
and the minus sign holds for models~7 and~8.
In both cases one may choose,
for instance,
real $f_1$ and $g_2$ while $g_1$ remains complex.
\begin{table}[ht!]
\centering
\begin{tabular}{c|cc}
& $H_1$ & $\left( H_2, H_3 \right)$
\\ \hline \\*[-4mm]
model~5 &
$\left( \overline{\mathbf{126}}, \mathbf{1} \right)$ &
$\left( \mathbf{10}, \mathbf{2} \right)$
\\ \hline \\*[-4mm]
model~6 &
$\left( \mathbf{10}, \mathbf{1} \right)$ &
$\left( \overline{\mathbf{126}}, \mathbf{2} \right)$
\\ \hline \\*[-4mm]
model~7 &
$\left( \overline{\mathbf{126}}, \mathbf{1} \right)$ &
$\left( \mathbf{120}, \mathbf{2} \right)$
\\ \hline \\*[-4mm]
model~8 &
$\left( \mathbf{10}, \mathbf{1} \right)$ &
$\left( \mathbf{120}, \mathbf{2} \right)$
\\ \hline
\end{tabular}
\caption{Minimal $SO(10)$ models with symmetry $O(2)$.
In each parenthesis,
the first number denotes the $SO(10)$ irrep
and the second number denotes the $O(2)$ irrep.
For symmetry $D_4$,
which is a subgroup of $O(2)$,
one should write $\mathbf{1}_{++}$ instead of $\mathbf{1}$.}
\label{table-D4models-2}
\end{table}

However,
the matrices~(\ref{vvipt}) have an accidental $U(1)_1$ continuous symmetry:
they are actually $O(2)$-symmetric.
The continuous group $O(2) = U(1)_1 \rtimes \Z_2$,
where $\Z_2$ is generated by $B$,
contains the discrete groups $D_n,\
\forall n \in \mathbb{N}$,
and therefore the matrices~(\ref{vvipt}) are invariant under any $D_n$,
in particular $D_4$ and $D_3 = S_3$.
In order to obtain a model that is invariant just under $D_4$
one may rely on the scalar potential;
some terms may be present in it
that break $O(2)$ down to its subgroup $D_4$.\footnote{If we had used for $H_1$
any other singlet of $D_4$ instead of $\mathbf{1}_{++}$,
then the full symmetry group of the ensuing matrices would have been
$U(1)_2 \times \left( U(1)_1 \rtimes \Z_2 \right) = U(1)_2 \times O(2)$.
Again,
this larger symmetry might be broken down to $D_4$
through the scalar potential.}

An alternative possibility to obtain a model that is just $D_4$-invariant,
and not also $O(2)$-invariant,
requires the use of four scalar multiplets,
by combining two distinct $D_4$ singlets, 
for example by adding to~(\ref{vvipt}) one further matrix
\be
\Gamma_4 = \Yukawa{0}{0}{0}{0}{f_2}{0}{0}{0}{f_2}.
\ee
The symmetry group of this set of four Yukawa-coupling matrices
will be $D_4$ and the mass matrices will not be block-diagonal.

\subsubsection{$D_4 = \left( \Z_2\times \Z_2 \right) \rtimes \Z_2$}

The minimal structure with symmetry $\Z_2 \times \Z_2$
is the single matrix~(\ref{Z2Z2-type}):
\be
\Gamma_0 = \Yukawa{f_1}{0}{0}{0}{f_2}{0}{0}{0}{f_3},
\ee
which is invariant under the $\Z_2 \times \Z_2$ group
$\left\{ 1,\ P_1,\ P_2,\ P_3 \right\}$,
where
\be
P_1 = \mathrm{diag} \left( +1,\ -1,\ -1 \right),
\quad
P_2 = \mathrm{diag} \left( -1,\ +1,\ -1 \right),
\quad
P_3 = \mathrm{diag} \left( -1,\ -1,\ +1 \right) = P_1 P_2. 
\ee
The automorphisms of $\Z_2 \times \Z_2$
form the $S_3$ group of the permutations of $P_1$,
$P_2$,
and $P_3$.
One can pick the $\Z_2$ subgroup of this $S_3$
which effects $P_2 \leftrightarrow P_3$.
This $\Z_2$ subgroup is generated by
the matrix $B$ of~(\ref{b-generator}).
Under that matrix,
$\Gamma_0$ transforms as
\be
B^T \Gamma_0 B = \mathrm{diag} \left( f_1,\ f_3\, e^{- 2 i \psi_2},\
f_2\, e^{2 i \psi_2} \right).
\ee
Therefore,
invariance of $\Gamma_0$ means
$f_3 = \pm f_2 \exp{\left( 2 i \psi_2 \right)}$
and implies $\left| f_2 \right| = \left| f_3 \right|$.
As before,
one may use the rephasing freedom to set $\psi_2 = 0$ and $f_2 = f_3$.

However,
$\Gamma_0 = \mathrm{diag} \left( f_1,\ f_2,\ f_2 \right)$ has a $U(1)$ symmetry,
given by the arbitrary rotation
between the second and third generations.
Besides,
$\Gamma_0$ by itself alone leads to diagonal fermion mass matrices,
hence no mixing.
Therefore we must accompany $\Gamma_0$ with a doublet of $D_4$.
We thereby reproduce the cases considered in the previous subsection,
albeit in a different weak basis.

\subsection{Models based on $\Z_4\, .\, \Z_2 = Q_4$}

The quaternion group $Q_4$,
which is of order eight,
is generated by $t_4$,
which satisfies $\left( t_4 \right)^4 = e$,
and a second generator $t$ satisfying $t^{-1} t_4 t = \left( t_4 \right)^{-1}$
but $t^2 \neq e$.
Still,
$t$ must be such that $t$ and $t_4$ do generate a finite group;
this is achieved if $t^2 = \left( t_4 \right)^2$,
since $\left( t_4 \right)^2$ generates the center $\Z_2$ of $Q_4$.
An explicit triplet representation of $Q_4$ through $SU(3)$ matrices is
\be
t_4 \to A_4 =
\left( \begin{array}{ccc}
1 & 0 & 0 \\ 0 & i & 0 \\ 0 & 0 & -i
\end{array} \right),
\quad
t \to C = \left(\begin{array}{ccc}
1 & 0 & 0 \\
0 & 0 & e^{i \psi} \\
0 & - e^{- i \psi} & 0 \\
\end{array} \right).
\label{c-generator}
\ee
The phase $\psi$ in $C$ is arbitrary.

The group $Q_4$ has four inequivalent singlet representations $\mathbf{1}_{pq}$,
where $p = \pm 1$ and $q = \pm 1$ just as in $D_4$,
with $t_4 \to p$ and $t \to q$.
The crucial difference between $D_4$ and $Q_4$ is that
$Q_4$ is a subgroup of $SU(2)$ but $D_4$ is not.
As a consequence,
the invariant $\mathbf{1}_{++}$ of $D_4$ lies in the symmetric part
of the product of two doublets,
while the invariant $\mathbf{1}_{++}$ of $Q_4$
is in the \emph{anti}symmetric part of the product.
This is seen in the matrices
\be
\Gamma^{(Q_4)}_{++} = \Yukawa{f}{0}{0}{0}{0}{g}{0}{-g}{0}, \quad
\Gamma^{(Q_4)}_{+-} = \Yukawa{0}{0}{0}{0}{0}{h}{0}{h}{0}, \quad
\Gamma^{(Q_4)}_{- \pm} = \Yukawa{0}{0}{0}{0}{l}{0}{0}{0}{\pm l\, e^{2 i \psi}},
\ee
which satisfy $A_4\, \Gamma^{(Q_4)}_{pq} A_4 = p\, \Gamma^{(Q_4)}_{pq}$
and $C^T\, \Gamma^{(Q_4)}_{pq}\, C = q\, \Gamma^{(Q_4)}_{pq}$.
One observes that $\Gamma^{(Q_4)}_{++}$,
which is $Q_4$-invariant,
is antisymmetric in the product of doublets
but symmetric in the product of singlets.
Now,
since the Yukawa-coupling matrices in an $SO(10)$ GUT
are always either symmetric or antisymmetric,
a scalar multiplet in the $\mathbf{1}_{++}$ of $Q_4$
will always couple through a Yukawa-coupling matrix $\Gamma^{(Q_4)}_{++}$
which features either $f=0$ or $g=0$.

This fact has drastic consequences,
namely,
all the $SO(10)$ Yukawa-coupling matrices with $Q_4$ symmetry
transform in a well-defined way under $U(1)_2$.
Therefore,
a set of $SO(10)$ Yukawa-coupling matrices with $Q_4$ symmetry alone,
unaccompanied by any $U(1)$ symmetries,
notably a symmetry of type $U(1)_2$,
is not possible.
Any $SO(10)$ model featuring $Q_4$ symmetry
must rely on the scalar potential to break
its accidental $U(1)_2$---and possibly also $U(1)_1$---symmetry.
This feature contrasts $Q_4$ with $D_4$ models.

\subsection{Models based on $A_4$ or $S_4$}
\label{sectionA4}

The $\Z_2 \times \Z_2$ group $\left\{ 1_{3\times3},\ P_1,\ P_2,\ P_3 \right\}$
has a group of automorphisms $S_3$,
formed by the permutations of $P_1$,
$P_2$,
and $P_3$.
The $\Z_3$ subgroup of this $S_3$ is generated by a $3\times3$ matrix $D$
such that $D^3 = 1_{3\times3}$ and $D^{-1} P_{1,2,3}\, D = P_{2,3,1}$.
One easily finds that that matrix $D$ is the one in~(\ref{D}).
Requiring $D^T \Gamma_0 D = e^{i \delta} \Gamma_0$ gives
\be
\label{biutp}
f_1 = e^{i \left( \delta - 2 \psi_1 \right)} f_2
= e^{- i \left( \delta + 2 \psi_1 + 2 \psi_2 \right)} f_3
\quad \mbox{and} \quad e^{3 i \delta} = 1.
\ee
One may thus define three matrices,
\bs
\label{A4-singlets}
\bea
\label{a40}
\Gamma_0^{(A_4)} &=& f_0\, \mathrm{diag} \left( 1,\ e^{2 i \psi_1},\
e^{2 i \left( \psi_1 + \psi_2 \right)} \right),
\\
\label{a41}
\Gamma_1^{(A_4)} &=& f_1\, \mathrm{diag} \left( 1,\ \omega^2\, e^{2 i \psi_1},\
\omega\, e^{2 i \left( \psi_1 + \psi_2 \right)} \right),
\\
\label{a42}
\Gamma_2^{(A_4)} &=& f_2\, \mathrm{diag} \left( 1,\ \omega\, e^{2 i \psi_1},\
\omega^2\, e^{2 i \left( \psi_1 + \psi_2 \right)} \right),
\eea
\es
corresponding to $e^{i \delta} = 1$,
$e^{i \delta} = \omega$,
and $e^{i \delta} = \omega^2$,
respectively.

The group $S_3$ is generated by $D$ in~(\ref{D})
together with $B$ in~(\ref{b-generator}).
Since $B^T \Gamma_0^{(A_4)} B = \Gamma_0^{(A_4)}$,
the matrix $\Gamma_0^{(A_4)}$ is $S_4$-invariant.
The matrices $\Gamma_1^{(A_4)}$ and $\Gamma_2^{(A_4)}$
transform into each other under the action of $B$,
provided $f_1 = f_2$.
Therefore,
the following is a doublet of $S_4$:
\be
\left\{ f_1\, \mathrm{diag} \left( 1,\ \omega^2\, e^{2 i \psi_1},\
\omega\, e^{2 i \left( \psi_1 + \psi_2 \right)} \right),\
f_1\, \mathrm{diag} \left( 1,\ \omega\, e^{2 i \psi_1},\
\omega^2\, e^{2 i \left( \psi_1 + \psi_2 \right)} \right)
\right\}.
\ee

It is clear that by using only Yukawa-coupling matrices $\Gamma_{0,1,2}^{(A_4)}$
one can only obtain diagonal fermion mass matrices,
impeding fermion mixing and making the model incompatible with phenomenology.
In order to allow for mixing one needs to include $A_4$/$S_4$ triplets.
There are two triplets of single-entry matrices,
a symmetric one and an antisymmetric one:
\be
\label{bvsit}
\Gamma_3 = \Yukawa{0}{g}{0}{\pm g}{0}{0}{0}{0}{0},
\quad
\Gamma_4 = g \Yukawa{0}{0}{0}{0}{0}{e^{i \left( \psi_1 + \psi_2 \right)}}{0}
{\pm e^{i \left( \psi_1 + \psi_2 \right)}}{0},
\quad
\Gamma_5 = g \Yukawa{0}{0}{\pm e^{i \psi_2}}{0}{0}{0}{e^{i \psi_2}}{0}{0}.
\ee
One of these triplets may accompany the matrix~(\ref{a40}).
The resulting models are given in table~\ref{table-A4modelsX}.
\begin{table}[ht!]
\centering
\begin{tabular}{c|cc}
& $H_1$ & $\left( H_2, H_3, H_4 \right)$
\\ \hline \\*[-4mm]
model 9 &
$\left( \overline{\mathbf{126}}, \mathbf{1} \right)$ &
$\left( \mathbf{10}, \mathbf{3} \right)$
\\ \hline \\*[-4mm]
model 10 &
$\left( \mathbf{10}, \mathbf{1} \right)$ &
$\left( \overline{\mathbf{126}}, \mathbf{3} \right)$
\\ \hline \\*[-4mm]
model 11 &
$\left( \overline{\mathbf{126}}, \mathbf{1} \right)$ &
$\left( \mathbf{120}, \mathbf{3}^\prime \right)$
\\ \hline \\*[-4mm]
model 12 &
$\left( \mathbf{10}, \mathbf{1} \right)$ &
$\left( \mathbf{120}, \mathbf{3}^\prime \right)$
\\ \hline
\end{tabular}
\caption{$SO(10)$ models with symmetry $S_4$.
In each parenthesis,
the first number denotes the $SO(10)$ irrep
and the second number denotes the $S_4$ irrep.}
\label{table-A4modelsX}
\end{table}
The corresponding Yukawa-coupling matrices are $\Gamma_0^{(A_4)}$
and $\Gamma_{3,4,5}$;
in the latter,
the plus sign holds for models~9 and~10
and the minus signs is for models~11 and~12.

The phases $\psi_1$ and $\psi_2$ may be rephased away while $f_0$ is made real;
the parameter $g$ in~(\ref{bvsit}) remains,
in general,
complex.

A model with the smaller symmetry $A_4$ requires an extra scalar multiplet,
coupling with either of the matrices~(\ref{a41}) or~(\ref{a42}).
Thus,
while a model with symmetry $S_4$ requires just four Yukawa-coupling matrices,
a model with the smaller symmetry $A_4$ needs at least five
Yukawa-coupling matrices.\footnote{This is reminiscent
of the situation with $O(2)$ and $D_4$,
studied in subsection~\ref{d4}.
A model with $O(2)$ symmetry needs only three Yukawa-coupling matrices,
a model with the smaller symmetry $D_4 \subset O(2)$ needs
at least four Yukawa-coupling matrices.}$^,$\footnote{Alternatively,
a model may have symmetry $S_4$ in its four Yukawa-coupling matrices
but that symmetry may be broken down to $A_4$ in the scalar potential.}

\subsection{Models based on $\Delta (54)$}

We have seen in subsection~\ref{Z3Z3}
that one may use the group $\Delta(27)$
for the Yukawa-coupling matrices in an $SO(10)$ GUT.
That subgroup of $SU(3)$
is generated by the matrices $A_3$ in~(\ref{a3-generator})
and $D$ in~(\ref{D}).
The group $\Delta(27)$ has nine triplet irreps $\mathbf{1}_{pq}$,
with $p, q \in \left\{ 0,\, 1,\ 2 \right\}$,
under which $A_3 \to \omega^p$ and $D \to \omega^q$.
Unfortunately,
though,
these singlet irreps cannot be realized in matrix form,
\textit{i.e.}\ there is no matrix $X$ such that
$A_3 X A_3 = \omega^p X$ and $D^T X D = \omega^q X$ simultaneously.
So,
one must realize $\Delta(27)$ solely through triplets
\be
\begin{array}{c}
\Gamma_1 = \Yukawa{f}{0}{0}{0}{0}{g}{0}{g}{0},
\quad
\Gamma_2 = \Yukawa{0}{0}{g\, e^{- i \psi_1}}{0}{f\, e^{2 i \psi_1}}{0}
{g\, e^{- i \psi_1}}{0}{0},
\\*[8mm]
\Gamma_3 = \Yukawa{0}{g\, e^{- i \left( \psi_1 + \psi_2 \right)}}{0}
{g\, e^{- i \left( \psi_1 + \psi_2 \right)}}{0}{0}
{0}{0}{f\, e^{2 i \left( \psi_1 + \psi_2 \right)}}.
\end{array}
\ee
These triplets actually are symmetric under a larger group than $\Delta(27)$,
namely $\Delta(54)$.

In order to construct a viable $\Delta(54)$-symmetric model
one must pick two triplets.
Minimal examples are given in table~\ref{table-Delta54}.
\begin{table}[ht!]
\centering
\begin{tabular}{c|cc}
& $\left( H_1, H_2, H_3 \right)$ & $\left( H_4, H_5, H_6 \right)$
\\ \hline \\*[-4mm]
model 13 &
$\left( \overline{\mathbf{126}}, \mathbf{3} \right)$ &
$\left( \mathbf{10}, \mathbf{3} \right)$
\\ \hline \\*[-4mm]
model 14 &
$\left( \mathbf{10}, \mathbf{3} \right)$ &
$\left( \mathbf{120}, \mathbf{3}^\prime \right)$
\\ \hline \\*[-4mm]
model 15 &
$\left( \overline{\mathbf{126}}, \mathbf{3} \right)$ &
$\left( \mathbf{120}, \mathbf{3}^\prime \right)$
\\ \hline
\end{tabular}
\caption{Minimal $SO(10)$ models with symmetry $\Delta(54)$.
In each parenthesis,
the first number denotes the $SO(10)$ irrep
and the second number denotes the $\Delta(54)$ irrep.}
\label{table-Delta54}
\end{table}
The corresponding Yukawa-coupling matrices are
\be
\label{uw1}
\begin{array}{c}
\Gamma_1 = \Yukawa{f}{0}{0}{0}{0}{g}{0}{g}{0}, \quad
\Gamma_2 = \Yukawa{0}{0}{g x^*}{0}{f x^2}{0}{g x^*}{0}{0}, \quad
\Gamma_3 = \Yukawa{0}{g x^* y^*}{0}{g x^* y^*}{0}{0}{0}{0}{f x^2 y^2},
\\*[8mm]
\Gamma_4 = \Yukawa{f^\prime}{0}{0}{0}{0}{g^\prime}{0}{g^\prime}{0}, \quad
\Gamma_5 = \Yukawa{0}{0}{g^\prime x^\ast}{0}{f^\prime x^2}{0}
{g^\prime x^\ast}{0}{0}, \quad
\Gamma_6 = \Yukawa{0}{g^\prime x^\ast y^\ast}{0}{g^\prime x^\ast y^\ast}{0}
{0}{0}{0}{f^\prime x^2 y^2},
\end{array}
\ee
for model~13,
and
\be
\label{uw2}
\begin{array}{c}
\Gamma_1 = \Yukawa{f}{0}{0}{0}{0}{g}{0}{g}{0}, \quad
\Gamma_2 = \Yukawa{0}{0}{g x^\ast}{0}{f x^2}{0}{g x^\ast}{0}{0}, \quad
\Gamma_3 = \Yukawa{0}{g x^\ast y^\ast}{0}{g x^\ast y^\ast}{0}{0}{0}{0}{f x^2 y^2},
\\*[8mm]
\Gamma_4 = \Yukawa{0}{0}{0}{0}{0}{g^\prime}{0}{-g^\prime}{0}, \quad
\Gamma_5 = \Yukawa{0}{0}{-g^\prime x^\ast}{0}{0}{0}{g^\prime x^\ast}{0}{0}, \quad
\Gamma_6 = \Yukawa{0}{g^\prime x^\ast y^\ast}{0}{-g^\prime x^\ast y^\ast}{0}{0}
{0}{0}{0},
\end{array}
\ee
for models~14 and~15,
where $x = \exp{\left( i \psi_1 \right)}$
and $y = \exp{\left( i \psi_2 \right)}$.

It is easy to check that,
through rephasings of the generations,
the phases $\psi_1$ and $\psi_2$ may be eliminated
while $f$ is rendered real in both~(\ref{uw1}) and~(\ref{uw2});
but,
$g$,
$g^\prime$,
and---in~(\ref{uw1})---$f^\prime$ will in general remain complex.

\subsection{Models based on $\Sigma (36)$}

The symmetry group $\Delta(54) \left/ \ \Z_3^\mathrm{center} \right.$
may be further enlarged,
to $\Sigma (36)$,
if we require the sets $\left\{ \Gamma_1,\ \Gamma_2,\ \Gamma_3 \right\}$
and $\left\{ \Gamma_4,\ \Gamma_5,\ \Gamma_6 \right\}$
of either~(\ref{uw1}) or~(\ref{uw2}) to be invariant under the action of
\be
\frac{i}{\sqrt{3}} \left(\begin{array}{ccc}
1 & e^{i \psi_1} & e^{i \left( \psi_1 + \psi_2 \right)} \\
e^{- i \psi_1} & \omega^2 & \omega e^{i \psi_2} \\
e^{- i \left( \psi_1 + \psi_2 \right)} & \omega e^{- i \psi_2} & \omega^2 \\
\end{array} \right).
\label{f-generator}
\ee
It is easy to check that this only happens when
\be
\label{buipd}
\mathrm{either} \quad
\frac{g}{f} = e^{i \left( 2 \psi_1 + \psi_2 \right)}\, \frac{- 1 + \sqrt{3}}{2}
\quad \mathrm{or} \quad
\frac{g}{f} = e^{i \left( 2 \psi_1 + \psi_2 \right)}\, \frac{- 1 - \sqrt{3}}{2}
\ee
for~(\ref{uw2});
for~(\ref{uw1}),
the condition~(\ref{buipd}) must apply and moreover
\be
\mathrm{either} \quad
\frac{g'}{f'} = e^{i \left( 2 \psi_1 + \psi_2 \right)}\, \frac{- 1 + \sqrt{3}}{2}
\quad \mathrm{or} \quad
\frac{g'}{f'} = e^{i \left( 2 \psi_1 + \psi_2 \right)}\, \frac{- 1 - \sqrt{3}}{2}
\ee
must also hold.

Just as in models based on $\Delta (54) \left/ \ \Z_3^\mathrm{center} \right.$,
the phases $\psi_1$ and $\psi_2$ may be rephased away
together with the phase of $f$.
Then,
$g$ and $f$ will be both real,
while $g^\prime$ and $f^\prime$ will be complex but have the same phase
(possibly apart from $\pi$).

There are only three discrete subgroups of $PSU(3)$
which contain $\Sigma(36)$ as a subgroup:
$\Sigma (72)$,
$\Sigma (216)$,
and $\Sigma (360)$.\footnote{\label{ludl} We thank Patrick Otto Ludl
for informing us about this fact.}
By looking at their generators~\cite{Grimus:2010ak},
one easily sees that the $\Sigma(36)$-invariant models
delineated above cannot be made invariant
under any larger discrete subgroup of $PSU(3)$.

\section{Discussion and conclusions} \label{conclusions}

Through the long history of model building within $SO(10)$ GUTs
equipped with flavour symmetry groups,
virtually all the studies have focused on specific (discrete) groups
and have studied their phenomenological consequences.
One might expect that,
the more scalars in different irreps of $SO(10)$ one would introduce,
the more complicated flavour symmetries one would be able to achieve
and the more elaborate Yukawa sectors one would construct,
with apparently limitless complexity.
In this paper we have shown that things are much more certain:
only a limited number of flavour symmetries may be achieved,
no matter how large the scalar sector of the $SO(10)$ GUT is.
We have given the full classification
of all possible flavour symmetry groups
that may be imposed on the Yukawa matrices,
for an arbitrary number of scalars in the $\mathbf{10}$,
$\overline{\mathbf{126}}$,
or $\mathbf{120}$ of $SO(10)$.
We have used methods from finite group theory
to identify all the possible non-Abelian discrete groups
whose Yukawa sector does not possess an accidental continuous symmetry.

We have also given examples of minimal models based on each discrete group.
Which of those examples might constitute truly viable phenomenological models,
\textit{viz.}\ able to fit the data,
remains yet to be studied.
In any case, we now know that
there exist no other essentially new possibilities beyond those that we have found.

One point that we have delegated to future studies
is the structure of the scalar sector for each model.
The scalar potential must be compatible with the required symmetry
and its minimization must lead to a vacuum that
breaks the symmetry just enough to produce a realistic fermion mixing,
but not so much that all predictive power gets lost.
This might not be easy to achieve:
scalar potentials with large symmetry groups
tend to have minima which break the symmetry only partially,
and the residual symmetries might render the fermion mass matrices
much too restrictive~\cite{no-go}.

\paragraph{Acknowledgements:} See footnote~\ref{ludl}.
This work was supported by the Portuguese
\textit{Fun\-da\-\c{c}\~{a}o para a Ci\^{e}ncia e a Tecnologia} (FCT)
under contracts UID/FIS/00777/2013 and CERN/
FIS-NUC/0010/2015,
which are partially funded through POCTI (FEDER),
COMPETE,
QREN
and the EU.
I.P.I.\ acknowledges funding from FCT
through the Investigator contract IF/00989/2014/CP1214/CT0004
under the IF2014 Program.

\end{document}